\documentclass[prl,aps,twocolumn,showpacs,preprintnumbers,amsmath,amssymb]{revtex4-2}

\usepackage{graphicx}
\usepackage{dcolumn}
\usepackage{bm}
\usepackage{color}
\usepackage{braket}
\usepackage{soul}
\usepackage{amsmath}
\usepackage{pdfpages}

\makeatletter
\define@key{Gin}{artifact}[]{}
\AtBeginDocument{\let\LS@rot\@undefined}
\makeatother

\begin{document}

\title{Unconventional Mixed-Parity Magnetism in Rare-Earth Tetraborides}
\author{Dong-Choon Ryu$^{1,2}$}
\email[dcrhyu@postech.ac.kr]{}
\author{Jae-Ho Han$^{3}$}
\author{Bongjae Kim$^{4}$}
\email[Co-corresponding: bongjae@knu.ac.kr]{}
\author{Chang-Jong Kang$^{1,2}$}
\email[Co-corresponding: cjkang87@cnu.ac.kr]{}

\affiliation{
$^1$Department of Physics, Chungnam National University, Daejeon 34134, Korea \\
$^2$Institute for Sciences of the Universe, Chungnam National University, Daejeon, 34134, Korea \\
$^3$Department of Physics and Chemistry, Korea Military Academy, Seoul 01805, Korea\\
$^4$Department of Physics, Kyungpook National University, Daegu, 41566, Korea
}
\date{\today}

\begin{abstract}
Altermagnetism has advanced the study of compensated magnets by revealing non-relativistic spin splitting,
traditionally classified into strictly even- or odd-parity spin textures.
Here, we establish component-resolved mixed-parity spin splitting
in a fully three-dimensional compensated magnet and identify its microscopic origin.
Using first-principles calculations, tight-binding and $\mathbf{k} \cdot \mathbf{p}$ models, along with spin-group symmetry analysis,
we demonstrate that the non-coplanar ground state of $\mathrm{TbB}_4$ enforces
odd-parity $p$- and $f$-wave-like textures for the in-plane spin components
while the out-of-plane component retains an even-parity $d$-wave altermagnetic character.
Crucially, the $p$-wave texture is driven not by relativistic spin-orbit coupling
but by a staggered emergent flux arising from the scalar spin chirality of the non-coplanar order.
This mixed-parity structure dictates distinct transport fingerprints,
including bulk non-relativistic Edelstein and spin Hall responses
that respond characteristically to reversal of the chirality domains.
Spin-orbit coupling additionally enables a Berry curvature dipole.
These results establish the rare-earth tetraborides as a robust platform
for engineering complex spin-charge conversion phenomena.
\end{abstract}
\maketitle


\begin{figure*}[t]
\includegraphics[width=7in]{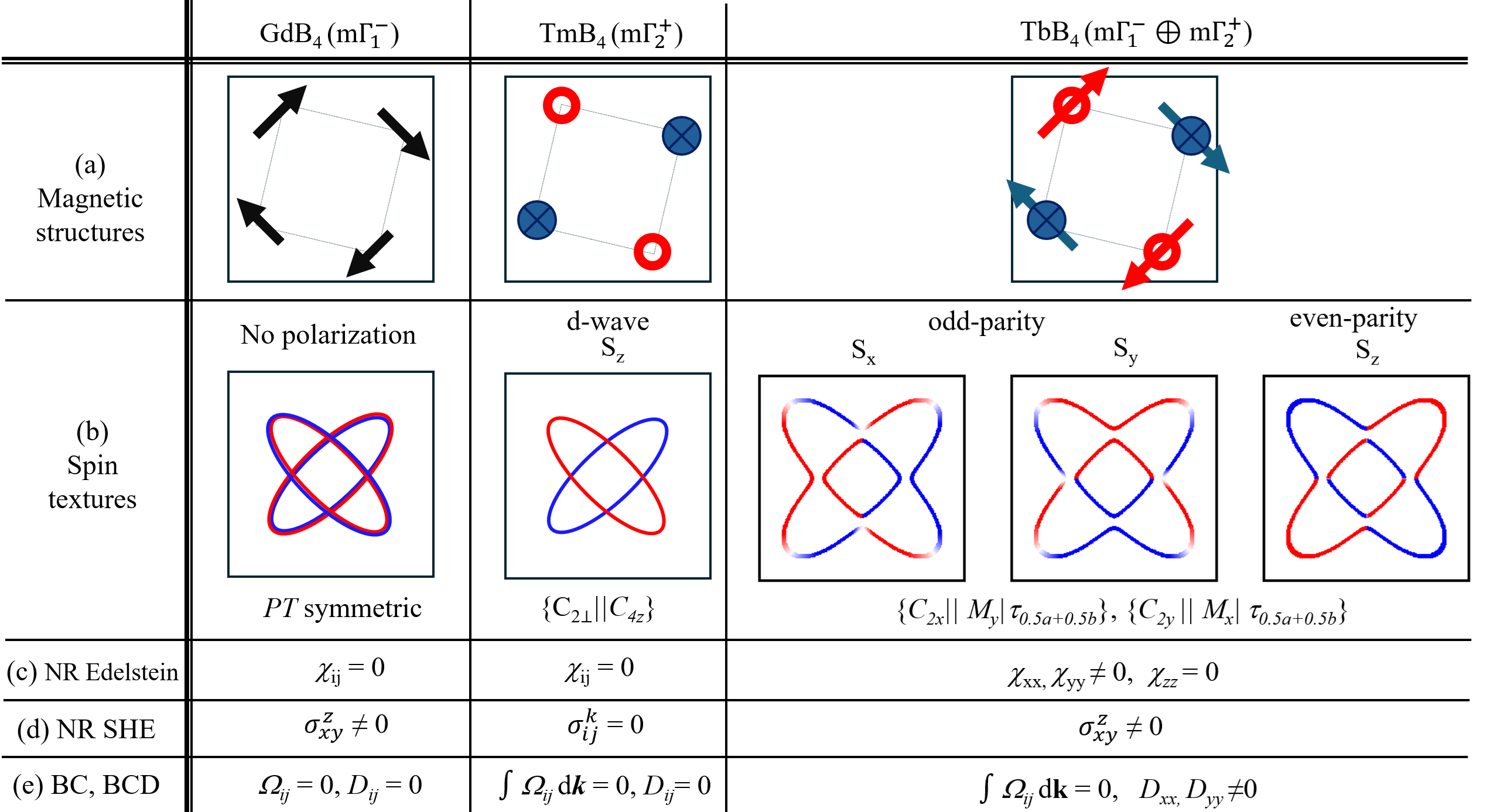}
\caption{(Color Online)
Comparative summary of the symmetry-driven phenomena
in $\mathrm{GdB}_4$, $\mathrm{TmB}_4$, and $\mathrm{TbB}_4$.
(a) Magnetic structures of the three compounds.
In $\mathrm{GdB}_4$, the preserved $PT$ symmetry completely suppresses
(b) the Fermi surface spin polarization,
(c) the non-relativistic Edelstein effect (NREE) tensor $\chi_{ij}$,
and (e) the local Berry curvature (BC),
but permits a nonzero (d) intrinsic ($T$-even)
non-relativistic spin Hall effect (NR SHE, $\sigma^{z}_{xy} \neq 0$).
$\mathrm{TmB}_4$ acts as a prototypical $d$-wave altermagnet with a robust spin-split Fermi surface;
however, its preserved spatial inversion alongside multiple (screw) rotational
and (glide) mirror symmetries enforces zero NREE, zero intrinsic ($T$-even) NR SHE,
zero net BC, and zero Berry curvature dipole (BCD).
In stark contrast, $\mathrm{TbB}_4$ hosts a unique mixture of odd-parity ($s_x$, $s_y$)
and even-parity ($s_z$) spin textures dictated by the
non-symmorphic spin-group symmetries
$\{C_{2x}\| M_{y}|\tau_{(a/2+b/2)}\}$ and $\{C_{2y}\| M_{x}|\tau_{(a/2+b/2)}\}$.
Crucially, the broken spatial inversion symmetry in $\mathrm{TbB}_4$ activates
a nonzero NREE for the in-plane diagonal components ($xx$ and $yy$),
and the even-parity $s_z$ texture supports
a finite NR SHE ($\sigma^{z}_{xy} \neq 0$)
for both the $T$-even and $T$-odd contributions,
while the $\{E\|M_{z}\}$ and $C_{4z}PT$ symmetries explicitly force
the NREE tensor $\chi_{zz}$ (without SOC)
and BCD $D_{zz}$ (with SOC) to be zero, respectively.
A finite BCD ($xx$ and $yy$ components) appears only with SOC.
Fermi surfaces in (b) are depicted around the $M$ point.
See the main text for further details.
}
\label{summary}
\end{figure*}

{\bf Introduction} --
Altermagnetism (AM) has recently emerged as a distinct class of magnetism \cite{Smejkal20, Smejkal22-1, Smejkal22-2,Smejkal22-3, Jungwirth25, Mazin22},
attracting growing research interest
as a platform for symmetry-driven spintronics \cite{Zhou25, Zhang25,Noh25,Liao24}.
AM combines features of both ferromagnetism (FM) and antiferromagnetism (AFM):
it exhibits spin splitting in the electronic structure, as in FM,
while retaining the compensated net magnetization characteristic of AFM.
This unusual coexistence of spin splitting and compensated magnetism distinguishes AM
from conventional magnetic phases.
A defining hallmark of AM is the momentum $\mathbf{k}$-dependent even-parity spin splitting,
$\varepsilon(s, \mathbf{k}) = \varepsilon(s, -\mathbf{k})$ (spin $s=\uparrow, \downarrow$),
which is typically characterized by \emph{d-}, \emph{g-}, or \emph{i-}wave spin textures.
Representative material candidates include RuO$_{2}$
\cite{Feng22, Fedchenko24, Noh25,Hernandez21,Shao21,Smejkal22-4},
MnTe \cite{Lee24,Amin24,Krempaský24, Osumi24,Hariki24,Takegami25},
and CrSb \cite{Ding24,Reimers24, Yang25, Biniskos25, Li25},
all of which feature collinear magnetic orders.
Recently, this paradigm has been extended to non-collinear or non-coplanar magnetic orders.
For instance, MnTe$_2$ exhibits an even-parity \emph{d}-wave spin texture constrained by orthogonal glide-mirror symmetries \cite{Zhu24}.

Parallel to these even-parity phases,
recent developments have further broadened the landscape to include odd-parity compensated magnets
\cite{hayami20, Hayami22, Brekke24, Song25,Zhuang25,Yamada2025,McNally2026,Meier26, zhang26}.
In such systems, non-collinear but coplanar magnetic orders drive a distinctive spin splitting satisfying
$\varepsilon(s,\mathbf{k})=\varepsilon(-s,-\mathbf{k})$,
which is typically protected by non-symmorphic spin-group symmetries
\cite{Yu25, Song25, Zhuang25, Liu22,Xiao24, Chen24,Jiang24}.
Such odd-parity spin textures are particularly important because they can generate a bulk non-relativistic Edelstein effect, \emph{i.e.}, current-induced spin accumulation without relying on spin-orbit coupling,
in contrast to the conventional Rashba-Edelstein mechanism \cite{Pari25,GonzalezHernandez24,Hu25,Chakraborty25,Neumann26}.
These findings establish odd-parity magnets as a distinct route to efficient spin-charge conversion, complementary to the charge-to-spin responses recently discussed in even-parity AMs \cite{Lai2025}.

Despite these advances, even- and odd-parity spin textures have mostly been treated as separate symmetry classes.
However, a natural bulk material in which these spin textures coexist in symmetry-distinct
spin channels--producing experimentally accessible transport fingerprints--remains elusive.
This motivates the search for compensated magnets hosting component-resolved mixed-parity spin splitting, where magnetoelectric and spin-transport responses normally associated with separate parity classes can emerge within a single magnetic phase.

Rare-earth tetraborides ($R$B$_4$) provide a natural material platform for this,
owing to their high-symmetry crystal symmetries and diverse, element-specific magnetic ground states
\cite{Blanco06, Ye17, Johnson24, Sim2016, Etourneau1979,Buschow1972,Choi2009}.
For instance, $\mathrm{PrB}_4$ has a ferromagnetic ground state \cite{Wigger05},
whereas $\mathrm{TmB}_4$ and $\mathrm{ErB}_4$ have collinear AFM order \cite{Ye17}.
$\mathrm{GdB}_4$ adopts a non-collinear but coplanar AFM structure
that breaks both inversion ($P$) and time-reversal ($T$) symmetries individually,
but preserves the combined $PT$ symmetry \cite{Blanco06}.
Recently, the magnetic structure of $\mathrm{TbB}_4$ has been resolved
and appears to be distinct from those of other tetraborides.
Originally interpreted as a chiral AFM order that breaks both $P$ and $T$ symmetries~\cite{Misawa2023},
it has been more accurately reassigned as a complex non-coplanar AFM state~\cite{Johnson24}.
In this configuration, the magnetic order of $\mathrm{TbB}_4$ is a linear combination ($m\Gamma_{1}^{-} \oplus m\Gamma_{2}^{+}$)
of the two distinct magnetic irreducible representations characteristic of $\mathrm{TmB}_4$ ($m\Gamma_{2}^{+}$) and $\mathrm{GdB}_4$ ($m\Gamma_{1}^{-}$).
Since each component possesses distinct parity and time-reversal characteristics,
the superposed magnetic structure in $\mathrm{TbB}_4$ yields a complex symmetry landscape,
enabling unconventional mixed-parity magnetism.

To explore this possibility,
we combine first-principles calculations with tight-binding and $\mathbf{k} \cdot \mathbf{p}$ models,
along with spin-group symmetry analysis, for the rare-earth tetraborides
$\mathrm{GdB}_4$, $\mathrm{TmB}_4$, and $\mathrm{TbB}_4$,
as summarized in Fig.~\ref{summary}.
We show that $\mathrm{GdB}_4$ is a $PT$-symmetric compensated antiferromagnet with no spin splitting
and $\mathrm{TmB}_4$ a $d$-wave altermagnet, whereas in $\mathrm{TbB}_4$
the non-coplanar order enforces a component-resolved mixed-parity spin splitting:
odd-parity ($p$- and $f$-wave-like) in-plane components coexist with
an even-parity ($d$-wave) out-of-plane component.
We identify the microscopic origin of this mixed-parity spin texture
and show that these distinct parity components give rise to complementary
transport fingerprints, namely non-relativistic Edelstein and spin Hall responses.
Together, these results establish $\mathrm{TbB}_4$ as a natural material platform
for realizing and probing mixed-parity magnetism.

{\bf Computational methods}  --
First-principles calculations were performed using
the Vienna \text{Ab initio} Simulation Package (VASP)
implemented with the projector augmented-wave (PAW)
pseudo-potential method \cite{VASP,GGA,PAW}.
To properly account for the strong electron correlation
and accurately capture the magnetic ground states of the $R$B$_{4}$ systems,
we adopt the GGA + $U$ scheme with $U$ = 6 eV and $J$ = 0.75 eV.
Further computational details are provided in
Sec.~I of the Supplemental Material (SM)~\cite{SM_ref}.

{\bf Magnetic structures and spin textures}  --
As illustrated in Fig.~\ref{summary}, $\mathrm{GdB}_4$ and $\mathrm{TmB}_4$
exemplify two representative AFM orders within the $R\mathrm{B}_4$ family.
Adopting their established magnetic structures
from previous works~\cite{Blanco06, Ye17, Johnson24},
we evaluate the electronic band structures and spin textures in the absence of spin-orbit coupling (SOC).
The resulting decoupling of spin and spatial degrees of freedom
enables a classification of the emergent spin textures based on spin-group symmetry.
Detailed electronic structures for $R\mathrm{B}_4$ ($R$ = $\mathrm{Gd}$, $\mathrm{Tm}$, $\mathrm{Tb}$)
are provided in Sec.~II of the SM~\cite{SM_ref}.

For $\mathrm{GdB}_4$, the magnetic moments of the Gd atoms lie within the in-plane,
forming a non-collinear coplanar AFM structure described by the $m\Gamma_{1}^{-}$ irreducible representation.
This order preserves the combined $PT$ symmetry, even though both $P$ and $T$ are individually broken.
Consequently, all bands remain at least Kramers spin degeneracy throughout the entire Brillouin zone,
thereby producing no spin splitting in the band structure (See the schematic Fermi surface in Fig.~\ref{summary}(b); for further details, see Sec.~III of the SM~\cite{SM_ref}).

For $\mathrm{TmB}_4$,
the magnetic moments of the Tm atoms are collinearly ordered along the $c$ axis,
corresponding to the $m\Gamma_2^+$ symmetry.
Here, the moments of two diagonally opposite Tm atoms are aligned in the same direction,
while the other two are oriented oppositely (see Fig.~\ref{summary}(a)).
This magnetic structure preserves inversion symmetry ($P$)
but breaks both the time-reversal ($T$) and combined $PT$ symmetries.
Due to the anisotropic crystal potential of the boron atoms,
which is associated with the spin-group symmetry $\{C_{2\perp}\|C_{4z}\}$
(where $C_{2\perp}$ is a $C_2$ rotation about an axis perpendicular to the spin),
the spin texture of $\mathrm{TmB}_4$ exhibits $d$-wave symmetry.
Thus, we identify $\mathrm{TmB}_4$ as the first altermagnetic candidate within the $R\mathrm{B}_4$ family.

\begin{figure}[t]
\includegraphics[width=3.5in]{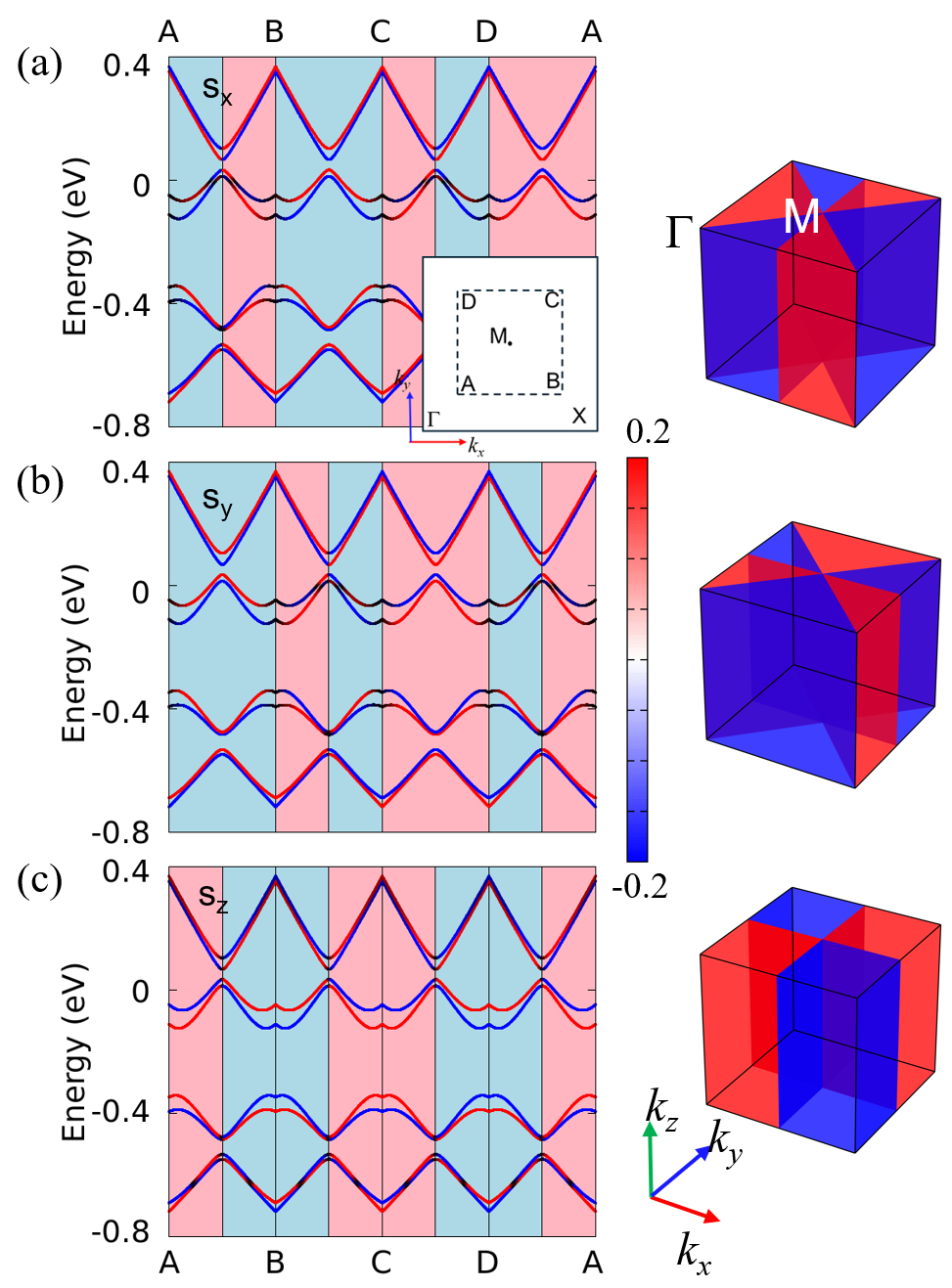}
\caption{(Color Online)
Component-wise spin splitting in the bulk band structures of $\mathrm{TbB}_4$.
(a)-(c) Spin-projected band structures evaluated at the $k_z$ = 0 plane
for the $s_x$, $s_y$, and $s_z$ components, respectively,
with the magnitude of spin polarization indicated by the color bar on the right.
The Fermi energy is set to zero.
The detailed \textbf{k}-path is illustrated in the inset of (a),
where the length ratio of AM to $\Gamma$M is set to 0.2.
The rightmost panels display schematic representations of the spin polarization distribution
for each component across the entire Brillouin zone (BZ).
Red and blue colors denote positive and negative spin polarizations, respectively.
}
\label{bs}
\end{figure}

The magnetic structure of $\mathrm{TbB}_4$ can be viewed as a linear superposition of those of $\mathrm{GdB}_4$ and $\mathrm{TmB}_4$,
characterized by the $m\Gamma_1^- \oplus m\Gamma_2^+$ symmetry~\cite{Johnson24}.
The out-of-plane magnetic moments comprise a collinear AFM order analogous to $\mathrm{TmB}_4$,
whereas the in-plane moments exhibit a non-collinear AFM order resembling $\mathrm{GdB}_4$.
However, the resulting spin texture is not a simple linear combination of the two.
In Fig.~\ref{bs}, the first column displays the spin-projected band structures along a $\mathbf{k}$-path with $k_z=0$, while the second column provides schematic representations of the spin polarization distribution for each component across the entire Brillouin zone (BZ). Notably, the out-of-plane spin component ($s_z$) exhibits an even-parity altermagnetic spin texture, whereas the in-plane spin components ($s_x$ and $s_y$) show odd-parity spin textures.

Now, we conduct a symmetry analysis of $\mathrm{TbB}_4$
(a full list of spin-group symmetry elements is provided in Appendix A).
The operations $\{C_{2x} \| M_{y} | \tau_{(a/2+b/2)}\}$ and $\{C_{2y} \| M_{x} | \tau_{(a/2+b/2)}\}$
within the $m\Gamma_1^- \oplus m\Gamma_2^+$ symmetry group yield the following relations in the band dispersion:
\begin{align}
 \varepsilon(\mathbf{s}_{\parallel}, \mathbf{k}) &= \varepsilon(-\mathbf{s}_{\parallel}, -\mathbf{k}), \nonumber \\
 \varepsilon(s_z, \mathbf{k}) &= \varepsilon(s_z, -\mathbf{k}),
\end{align}
where $\mathbf{s}_{\parallel} = (s_x, s_y)$ denotes the in-plane spin components
(see Appendix B for details).
These symmetry-enforced constraints strictly dictate the mixed-parity spin textures in $\mathrm{TbB}_4$,
characterized by odd-parity $s_x$ and $s_y$ components alongside an even-parity $s_z$ component,
a situation distinct from previously reported $p$-wave magnetism~\cite{Brekke24,hellenes2024},
in which in-plane spin polarization is forbidden.
The general symmetry criterion for two odd-parity in-plane and one even-parity out-of-plane components
is the presence of the spin-group element $\{C_{2z}\|P\}$
together with the absence of any spin-only element $\{I_s C_{2n}\|E\}$.
The latter condition requires a noncoplanar order,
since any coplanar order possesses the spin-only element $\{I_s C_{2n}\|E\}$
with $n$ the normal to the moment plane,
which forces the spin components perpendicular to $n$ to be even \cite{luo26}.

However, a symmetry classification only states which spin textures are symmetry-allowed.
What sets the mixture of $p$- and $f$-wave components in the in-plane polarization,
and why it appears as a $p$-wave-like inner and an $f$-wave-like outer Fermi pocket,
are questions that require a microscopic model.
We address them below and establish $\mathrm{TbB}_4$ as the first metallic realization
of this form of mixed-parity magnetism with a definite microscopic origin.

{\bf Microscopic origin} --
To identify the origin of the $p/f$-wave mixture of the in-plane polarization,
we construct a tight-binding model (see Sec.~IV of the SM \cite{SM_ref} for details).
We consider conduction electrons on the Shastry-Sutherland lattice
with the nearest-neighbor, dimer, and next-nearest-neighbor hoppings,
coupled on each site to the non-coplanar Tb moments via an $s$--$f$ exchange coupling $J_{sf}$.
In addition, we introduce a staggered complex hopping $\pm i\lambda$ on the nearest-neighbor bonds,
which arises from the scalar spin chirality $\chi_{ijk}=\bm S_i \cdot (\bm S_j \times \bm S_k)$
carried by the non-coplanar magnetic moments:
an electron hopping around a triangle of non-coplanar moments
acquires a Berry phase that is an odd function of the scalar spin chirality,
with a sign that alternates between glide-related plaquettes.
Downfolding onto an effective two-band model near $M$ gives
$\mathcal H_{\rm eff}(\mathbf q)=\varepsilon_0(\mathbf q)\sigma_0
+\mathbf h(\mathbf q)\cdot\boldsymbol\sigma$
with, to cubic order
(see Appendix C),
\begin{align}
\mathbf h_\parallel(\mathbf q) &= \alpha \lambda J_{sf} a\,(q_x,-q_y)\nonumber\\
&\quad + J_{sf} a^3 \big[c_1(q_x^3,-q_y^3) + c_2(q_xq_y^2,-q_x^2q_y)\big],\nonumber\\
h_z(\mathbf q) &= \gamma J_{sf} a^2\,q_xq_y ,
\label{eq:heff}
\end{align}
where $\alpha$, $c_1$, $c_2$, and $\gamma$ are constant coefficients.
The relative signs within each pair of components in $\mathbf h_\parallel$
are fixed by $\{I_s C_{4z}\|C_{4z}^{-1}\}$
and make the linear term an anti-vortex (winding number $-1$).
The two in-plane harmonics,
the linear $p$-wave term and the cubic $f$-wave term in Eq.~(\ref{eq:heff}),
have distinct origins.
In the tight-binding model, $c_1$ is proportional to the staggered complex hopping $\lambda$
(see Sec.~V of the SM \cite{SM_ref} for details).
Consequently, at $\lambda = 0$
both the linear $p$-wave term and $c_1$ vanish.
Only the mixed cubic term $c_2$ survives,
generated by the dimer hopping acting on the non-coplanar exchange field,
and produces the $f$-wave-like texture.
A finite $\lambda$ generates the $p$-wave anti-vortex of the inner pocket,
with amplitude proportional to $\lambda$.
For small $q$, the linear term dominates and gives a $p$-wave texture,
whereas for large $q$, the cubic terms dominate and give an $f$-wave texture.
This is consistent with the DFT result in Fig.~\ref{summary}(b)
(see also SFIG.~8 of the SM for the tight-binding spin texture),
where the in-plane spin texture is $p$-wave on the inner pocket and $f$-wave on the outer pocket,
and the out-of-plane component is $d$-wave.
Finally, the same classification holds on every $k_z$ plane,
with coefficients in Eq.~(\ref{eq:heff}) that are even functions of $k_z$
(see Secs.~VI and VII of the SM~\cite{SM_ref}).
Next, we investigate the physical consequences and observable quantities induced by this unique coexistence of even- and odd-parity spin textures in $\mathrm{TbB}_4$.

\begin{figure}[t]
\includegraphics[width=3.5in]{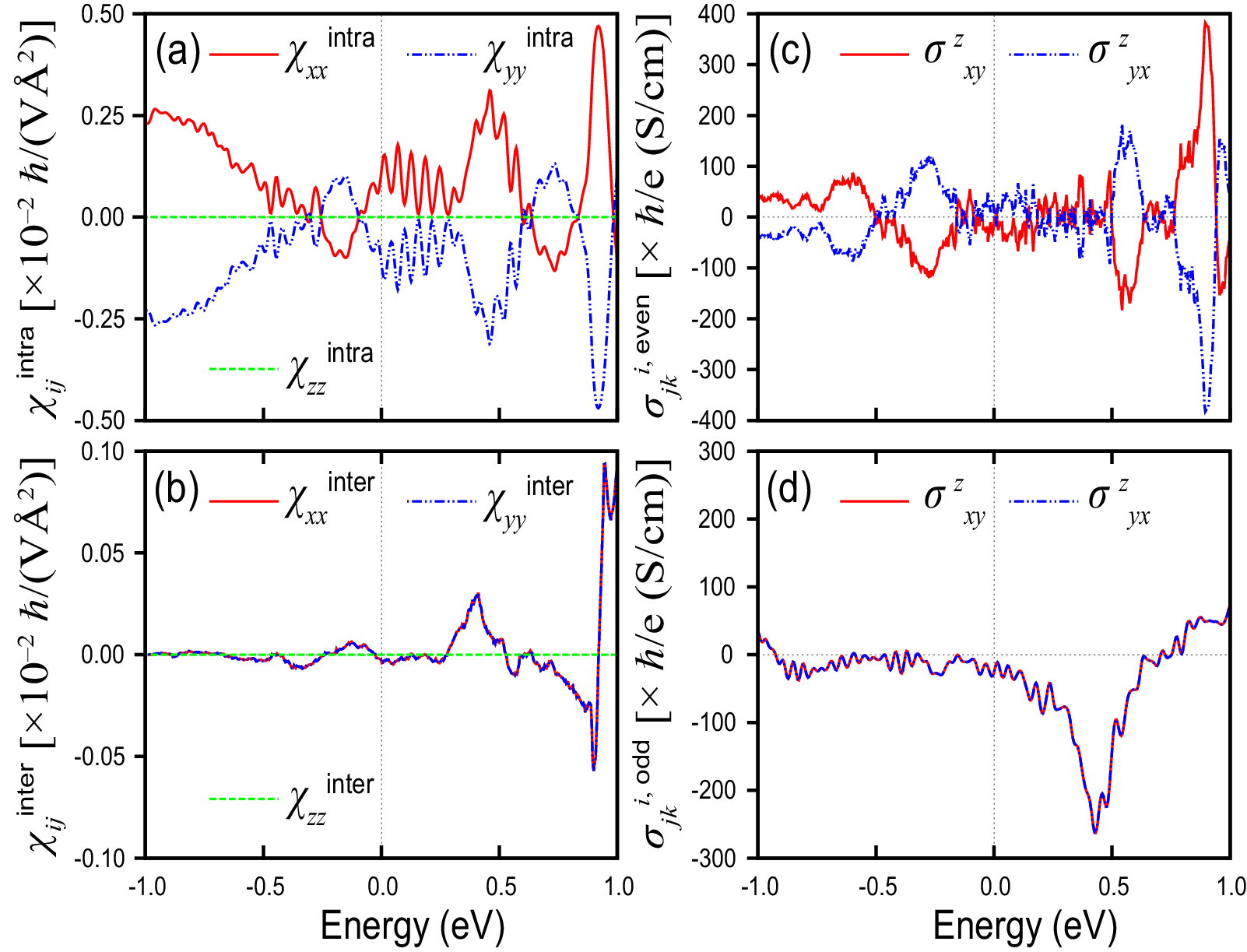}
\caption{(Color Online)
Non-relativistic Edelstein response tensors
(a) $\chi^{\text{intra}}_{ij}$ and (b) $\chi^{\text{inter}}_{ij}$
as functions of energy.
Due to the $\{I_s C_{4z}\|C_{4z}^{-1}\}$ symmetry,
the relations
$\chi^{\text{intra}}_{xx} = -\chi^{\text{intra}}_{yy}$ and
$\chi^{\text{inter}}_{xx} = \chi^{\text{inter}}_{yy}$ are strictly satisfied.
Non-relativistic spin Hall conductivities
$\sigma^{\text{z}}_{xy}$ and $\sigma^{\text{z}}_{yx}$
for the (c) $T$-even (intrinsic) and (d) $T$-odd (spin-splitter) components.
The same antiunitary element enforces
$\sigma^{z,{\rm even}}_{xy} = -\sigma^{z,{\rm even}}_{yx}$
and $\sigma^{z,{\rm odd}}_{xy}=\sigma^{z,{\rm odd}}_{yx}$.
The scattering rate is $\Gamma = 0.03$ eV
for both the Edelstein tensors and the spin Hall conductivities.
The Fermi energy is set to zero.
}
\label{edel}
\end{figure}

{\bf Non-relativistic Edelstein effect} --
The non-relativistic Edelstein effect (NREE) is a key transport fingerprint of odd-parity spin textures,
reflecting electric-field-induced spin accumulation without relying on SOC
\cite{GonzalezHernandez24,Hu25,Chakraborty25}.
Since $\mathrm{TbB}_4$ hosts odd-parity in-plane spin components,
a finite bulk NREE provides a direct probe of the odd-parity sector within its mixed-parity spin texture.
To quantify this, we evaluate the Edelstein tensor $\chi_{ij}$,
which describes spin accumulation $S_i$ induced by an external electric field $E_j$
via the relation $S_i = \chi_{ij} E_j$ ($i, j \in \{x, y, z\}$).
Following linear response theory~\cite{Li15, Hu25}, the total Edelstein tensor can be
decomposed into intraband ($T$-even) and interband ($T$-odd) contributions:
$\chi = \chi^{\mathrm{intra}} + \chi^{\mathrm{inter}}$
(see Appendix D for the Kubo expressions and the symmetry analysis).

From the spin-group symmetry analysis,
the only symmetry-allowed components are the in-plane diagonal elements $\chi_{xx}$ and $\chi_{yy}$.
As expected, our first-principles calculations yield finite profiles for both $\chi_{xx}$ and $\chi_{yy}$.
In Figs.~\ref{edel} (a) and (b), we plot these components alongside $\chi_{zz}$ as functions of energy.
Note that $\chi_{zz}=0$ is identically maintained, and the symmetry-enforced relations for the intra- and inter-band contributions are strictly satisfied.

{\bf Non-relativistic spin Hall effect} --
While the non-relativistic Edelstein response probes the odd-parity in-plane sector of $\mathrm{TbB}_4$,
the mixed-parity character of this system also requires a transport signature
associated with the even-parity out-of-plane spin texture.
The intrinsic ($T$-even) spin Hall
and spin-splitter ($T$-odd) effects,
both non-relativistic, provide such a complementary fingerprint
(see Appendix E for the Kubo expressions and the symmetry analysis):
the $T$-even (intrinsic) spin Hall conductivity $\sigma^{i,{\rm even}}_{jk}$
arises from the spin Berry curvature (Eq.~(\ref{eq:she_even})),
whereas the $T$-odd (spin-splitter) conductivity $\sigma^{i,{\rm odd}}_{jk}$
is generated directly by the $d$-wave spin splitting (Eq.~(\ref{eq:she_odd})).
Here $i$ and $j$ in $\sigma^i_{jk}$ denote the spin-polarization and current directions, respectively,
and $k$ the electric-field direction.

Based on symmetry arguments,
both $T$-even and $T$-odd contributions vanish for the odd-parity spin components.
For the even-parity component, only $\sigma^{z}_{xy}$ is allowed
and finite for both contributions,
with $\sigma^{z,{\rm even}}_{xy} = -\sigma^{z,{\rm even}}_{yx}$ and
$\sigma^{z,{\rm odd}}_{xy} = +\sigma^{z,{\rm odd}}_{yx}$.
The calculated conductivities in Figs.~\ref{edel}(c) and (d) confirm these relations.
The two contributions are further distinguished
by their behavior under reversal of the magnetic domain:
the $T$-odd part reverses with the scalar spin chirality,
whereas the $T$-even part does not (Sec.~XI of the SM~\cite{SM_ref}).

{\bf Effects of SOC} --
When SOC is taken into account, the symmetry analysis must transition to
the magnetic space group framework.
The mixed-parity spin texture is robust against SOC:
the in-plane $p/f$-wave mixture and out-of-plane $d$-wave textures
are preserved (Sec.~VIII of the SM~\cite{SM_ref}).
For the Edelstein response, SOC allows $\chi_{zz}$ in principle,
but $C_{4z}PT$ still enforces $\chi^{\rm intra}_{zz} = 0$
and leaves only a finite interband component $\chi^{\rm inter}_{zz}$
(see Sec.~IX of the SM~\cite{SM_ref} for details).
The most prominent effect of SOC concerns the Berry curvature.
Without SOC, the element $\{C_{2z}\|P\}$ acts as an inversion on the Berry curvature
and forces the Berry curvature dipole (BCD) to vanish.
With SOC, this element is removed.
The net Berry curvature still vanishes, as without SOC,
now enforced by the magnetic rotational symmetries
$C_{2x}\tau_{(a/2+b/2)}$, $C_{2y}\tau_{(a/2+b/2)}$, and $C_{2z}$,
but a finite BCD of the form $D_{xx} = -D_{yy}$, $D_{zz} = 0$
becomes symmetry-allowed and is indeed obtained (Sec.~X of the SM~\cite{SM_ref}),
giving rise to a nonlinear Hall effect in $\mathrm{TbB}_4$.

{\bf Concluding Remarks} --
We have demonstrated that component-resolved mixed-parity spin splitting can be realized in a fully three-dimensional compensated magnet, establishing rare-earth tetraborides as a natural platform for symmetry-engineered spin textures.
Within this family,
$\mathrm{GdB}_4$ is a $PT$-symmetric antiferromagnet with spin-degenerate bands
and $\mathrm{TmB}_4$ a $d$-wave altermagnet,
whereas $\mathrm{TbB}_4$, whose non-coplanar order combines
the $\mathrm{GdB}_4$- and $\mathrm{TmB}_4$-type magnetic irreducible representations,
hosts a momentum-space spin texture that is not a simple superposition of the two.
Instead, the nonsymmorphic spin-group symmetries of $\mathrm{TbB}_4$ enforce odd-parity
$p$- and $f$-wave-like textures for the in-plane spin components $s_x$ and $s_y$,
while the out-of-plane component $s_z$ retains an even-parity $d$-wave altermagnetic character.
Beyond this symmetry classification,
we have identified the microscopic origin of the coexistence of $p$- and $f$-wave-like textures.
The scalar spin chirality of the non-coplanar order generates a staggered emergent flux,
which is solely responsible for the linear $p$-wave anti-vortex of the inner Fermi pocket,
whereas the $f$-wave texture of the outer pocket arises from the dimer hopping
acting on the non-coplanar exchange field.
Symmetry thus fixes the parity class,
while the scalar spin chirality supplies the amplitude
and sign of the leading odd-parity term.

The coexistence of these mixed-parity sectors leads to distinct transport fingerprints:
the odd-parity in-plane texture generates a non-relativistic Edelstein response,
and the even-parity out-of-plane texture supports
both an intrinsic ($T$-even) and a spin-splitter ($T$-odd) non-relativistic
spin Hall conductivity.
Their behavior under reversal of the magnetic domain
---the intraband Edelstein and spin-splitter responses reverse
upon reversal of the out-of-plane local moments,
whereas the interband Edelstein and intrinsic spin Hall responses do not---
provides a direct experimental test of the proposed mechanism.
In the presence of SOC, $\mathrm{TbB}_4$ further develops a symmetry-allowed Berry curvature dipole,
providing an additional route to nonlinear Hall phenomena~\cite{sodemann15}.
These results establish $\mathrm{TbB}_4$ as the first metallic realization
of component-resolved mixed-parity magnetism with a definite microscopic origin,
and suggest rare-earth tetraborides and related Shastry-Sutherland magnets
as a materials family for non-relativistic spin-charge conversion in compensated magnets.
More generally, the symmetry criterion identified here is naturally met by superposing
a $P$-even and a $P$-odd magnetic irreducible representation on a nonsymmorphic lattice,
as in $\mathrm{TbB}_4$, which provides a route to further realizations.

\textit{Note added.}---
During the preparation of this manuscript we became aware of related theoretical works.
Refs.~\cite{Zhuang26,Yu26} obtain mixed-parity spin splitting in collinear magnets,
within a single spin component, through orbital order or external driving,
whereas our work establishes that in $\mathrm{TbB}_4$
the parity is definite for each spin component
and arises intrinsically from the non-coplanar order.
Ref.~\cite{luo26} establishes a general symmetry classification of mixed-parity magnets;
in that language $\mathrm{TbB}_4$ belongs to the type-III class,
for which the only candidate identified there, FeSb$_2$O$_4$,
has a symmetry-forbidden Edelstein response.
Our work, carried out independently,
goes beyond the classification by identifying the microscopic origin of the odd-parity components,
the chirality-induced staggered flux, and by providing a metallic realization
with finite Edelstein and spin Hall responses.
Ref.~\cite{shen26} considers umbrella-type orders
in which odd and even harmonics mix within a single spin component;
in our work the parity is instead definite for each spin component
and symmetry-enforced on every $k_z$ plane.

\textit{Acknowledgment}---
We are grateful to Igor Mazin, Hyun-Woo Lee, Yong Baek Kim, Hae-Young Kee, Kyoung-Whan Kim,
and Hosub Jin for useful discussions.
D.-C. R. and C.-J. K. were supported by
the National Research Foundation of Korea (NRF) (Grant No. 2022R1C1C1008200) and
the KISTI Supercomputing Center (Project No. KSC-2025-CRE-0461).
D.-C. R. was supported by NRF (Grant No. RS-2023-00274550).
C.-J. K. was also supported by NRF Grant funded by the Korean Government (MOE).
B.K. acknowledges support by NRF Grants
(No. RS-2024-00401881, No. RS-2026-25472078, and No. RS-2022-NR068223)
and KISTI supercomputing Center (Project No. KSC-2023-CRE-0413).
This research was supported by Global - Learning \& Academic research institution for Master’s·PhD students,
and Postdocs (G-LAMP) Program of NRF grant funded by the Ministry of Education (No. RS-2025-25442707).
This research was also supported by “Creation of the quantum information science R\&D ecosystem
(based on human resources)” through the NRF funded by the Korean government
(MSIT; Grant No. RS-2023-00256050).
We acknowledge the hospitality of the Aspen Center for Physics
during the 2026 Aspen Winter Conference
(Altermagnetism and Unconventional Magnetic Orders in Quantum Materials),
and of APCTP (Pohang, Korea) during the conference [APCTP-2025-C01],
where fruitful discussions greatly benefited this work.

\nocite{Heiba85}
\bibliography{cleaned.bib}

\section{End Matter}

\emph{Appendix A: Spin-space group symmetry elements in $\mathrm{TbB}_4$}---
Based on the spin-space group (SSG) analysis, the non-coplanar magnetic ground state of $\mathrm{TbB}_4$ comprises 16 symmetry operations. These elements are expressed using the generalized Seitz notation $\{R_s || R_r | \mathbf{t}\}$, incorporating Schoenflies crystallographic symbols:
\begin{align*}
    &\{E\|E|\tau_{(0,0,0)}\}, \quad \{E\|M_z|\tau_{(0,0,0)}\}, \\
    &\{C_{2z}\|C_{2z}|\tau_{(0,0,0)}\}, \quad \{C_{2z}\|P|\tau_{(0,0,0)}\}, \\
    &\{C_{2x}\|M_y|\tau_{(a/2+b/2)}\}, \quad \{C_{2x}\|C_{2x}|\tau_{(a/2+b/2)}\}, \\
    &\{C_{2y}\|M_x|\tau_{(a/2+b/2)}\}, \quad \{C_{2y}\|C_{2y}|\tau_{(a/2+b/2)}\}, \\
    &\{I_{s}C_{2(1\bar{1}0)}\|M_{110}|\tau_{(a/2+b/2)}\}, \quad \{I_{s}C_{2(1\bar{1}0)}\|C_{2(110)}|\tau_{(a/2+b/2)}\}, \\
    & \{I_{s}C_{2(110)}\|M_{1\bar{1}0}|\tau_{(a/2+b/2)}\}, \quad \{I_{s}C_{2(110)}\|C_{2(1\bar{1}0)}|\tau_{(a/2+b/2)}\}, \\
    &\{I_{s}C_{4z}\|C_{4z}^{-1}|\tau_{(0,0,0)}\}, \quad \{I_{s}C_{4z}\|S_{4z}^{-1}|\tau_{(0,0,0)}\}, \\
    &\{I_{s}C_{4z}^{-1}\|C_{4z}|\tau_{(0,0,0)}\}, \quad \{I_{s}C_{4z}^{-1}\|S_{4z}|\tau_{(0,0,0)}\}.
\end{align*}
In this notation, $R_s$ acts exclusively on the spin space, $R_r$ acts on the real-space atomic coordinates, and $\mathbf{t}$ denotes the fractional translation vector (e.g., $\tau_{(a/2+b/2)}$ indicates an in-plane translation by half a lattice parameter along both the $a$ and $b$ axes). The symmetry operations ($R_s$ and $R_r$) are denoted using the Schoenflies nomenclature:
\begin{itemize}
    \item $E$ and $P$: The identity and spatial inversion operations, respectively.
    \item $C_n$: An $n$-fold proper rotation.
    \item $M$: A mirror reflection across a plane.
    \item $S_4$: A 4-fold improper rotation (roto-reflection).
    \item $I_{s}$: Spin-only inversion.
\end{itemize}
The subscripts specify the Cartesian or crystallographic direction of the rotation axis or the normal to the mirror plane. The superscript $-1$ denotes the inverse of the operation (e.g., $C_{4z}^{-1}$ represents a $-90^\circ$ rotation).

\emph{Appendix B: Symmetry analysis for mixed-parity spin splitting
and spin degeneracy in $\mathrm{TbB}_4$}---
The non-symmorphic spin group symmetry elements
$\{C_{2x} \| M_{y} | \tau_{(a/2+b/2)}\}$ and $\{C_{2y} \| M_{x} | \tau_{(a/2+b/2)}\}$
impose the following constraints on the spin- and momentum-dependent
band dispersion $\varepsilon(\mathbf{s}, \mathbf{k})$, respectively:
\begin{align}
\varepsilon(s_x, s_y, s_z, k_x, k_y, k_z) &= \varepsilon(s_x, -s_y, -s_z, k_x, -k_y, k_z), \label{eq4} \\
\varepsilon(s_x, s_y, s_z, k_x, k_y, k_z) &= \varepsilon(-s_x, s_y, -s_z, -k_x, k_y, k_z). \label{eq5}
\end{align}
Combining these two relations yields:
\begin{align}
\varepsilon(s_x, s_y, s_z, k_x, k_y, k_z) &= \varepsilon(-s_x, -s_y, s_z, -k_x, -k_y, k_z). \nonumber
\end{align}
On each fixed $k_z$ plane, the above relation reduces to:
\begin{align}
\varepsilon(\mathbf{s}_{\parallel}, \mathbf{k}_{\parallel}) &= \varepsilon(-\mathbf{s}_{\parallel}, -\mathbf{k}_{\parallel}), \nonumber \\
\varepsilon(s_z, \mathbf{k}_{\parallel}) &= \varepsilon(s_z, -\mathbf{k}_{\parallel}),
\end{align}
where $\mathbf{s}_{\parallel} = (s_x, s_y)$ and $\mathbf{k}_{\parallel} = (k_x, k_y)$.
Finally, incorporating the spin-group symmetry $\{E\|M_z\}$ generalizes
these constraints to the full 3D momentum space:
\begin{align}
\varepsilon(\mathbf{s}_{\parallel}, \mathbf{k}) &= \varepsilon(-\mathbf{s}_{\parallel}, -\mathbf{k}), \nonumber \\
\varepsilon(s_z, \mathbf{k}) &= \varepsilon(s_z, -\mathbf{k}).
\end{align}
These symmetry-enforced constraints robustly explain the mixed-parity spin textures in $\mathrm{TbB}_4$,
which are characterized by odd-parity $s_x$ and $s_y$ components alongside an even-parity $s_z$ component.

Furthermore, when $k_x = 0$ or $\pi/a$ (where $-k_x \equiv k_x$),
Eq.~(\ref{eq5}) simplifies to
$\varepsilon(s_x, s_y, s_z, \mathbf{k}) = \varepsilon(-s_x, s_y, -s_z, \mathbf{k})$.
This indicates that states with opposite $s_x$ and $s_z$ spin components share the same energy dispersion,
thereby enforcing the exact spin degeneracy (i.e., vanishing spin polarization) of $s_x$ and $s_z$
on the $k_x = 0$ and $\pi/a$ planes.
In the same manner, when $k_y = 0$ or $\pi/a$, Eq.~(\ref{eq4}) dictates that
$\varepsilon(s_x, s_y, s_z, \mathbf{k}) = \varepsilon(s_x, -s_y, -s_z, \mathbf{k})$,
resulting in the spin degeneracy of $s_y$ and $s_z$ on the $k_y = 0$ and $\pi/a$ planes.

\emph{Appendix C: Effective $\mathbf{k} \cdot \mathbf{p}$ Hamiltonian and momentum-dependent spin textures near the $M$ point}---
Let $\mathbf{q} = (q_x, q_y, 0)$ be the small momentum deviation
from the $M$ point [$\mathbf{k} = (\pi/a, \pi/a, 0)$],
and write the effective two-band Hamiltonian as
$\mathcal{H}_{\text{eff}}(\mathbf{q})
= \varepsilon_0(\mathbf{q})\sigma_0 + \mathbf{h}(\mathbf{q}) \cdot \boldsymbol{\sigma}$,
where $\sigma_0$ is the $2 \times 2$ identity matrix, $\boldsymbol{\sigma}$ are the Pauli matrices, and $\mathbf{h}(\mathbf{q})$ is the effective spin-splitting field.
The nonsymmorphic spin-group symmetries
$\{C_{2x} \| M_{y} | \tau_{(a/2+b/2)}\}$ and $\{C_{2y} \| M_{x} | \tau_{(a/2+b/2)}\}$
strictly constrain the components of $\mathbf{h}(\mathbf{q})$
up to the third order in $\mathbf{q}$ as follows:
\begin{align}
h_x(\mathbf{q}) &= \alpha_1 q_x + \beta_1 q_x^3 + \gamma_1 q_x q_y^2, \nonumber \\
h_y(\mathbf{q}) &= \alpha_2 q_y + \beta_2 q_y^3 + \gamma_2 q_x^2 q_y, \nonumber \\
h_z(\mathbf{q}) &= \delta q_x q_y,
\label{eq:kp_model}
\end{align}
where $\alpha_1$, $\alpha_2$, $\beta_1$, $\beta_2$, $\gamma_1$, $\gamma_2$, and $\delta$
are constant coefficients.
The relative signs within each pair are fixed
by the antiunitary element $\{I_s C_{4z}\|C_{4z}^{-1}\}$,
whose spin part rotates $(s_x, s_y)$ by $-90^\circ$
while its momentum part rotates $\mathbf{q}$ by $+90^\circ$,
requiring $h_x(q_x, q_y) = h_y(q_y, -q_x)$.
Substituting Eq.~(\ref{eq:kp_model}) gives
$\alpha_2 = -\alpha_1$, $\beta_2 = -\beta_1$, and $\gamma_2 = -\gamma_1$,
so that each pair carries a single independent amplitude
and the leading in-plane field $\mathbf{h}_\parallel^{(1)}\propto(q_x,-q_y)$
is an anti-vortex of winding number $-1$ rather than a vortex.
In the representation-theoretic language,
$(s_x, s_y)$ and $s_z$ transform as the $E_u$ and $B_{2g}$ irreducible representations
of the effective point group $D_{4h}$ acting on $\mathbf{k}$,
and the same form holds on every $k_z$ plane with coefficients that are even in $k_z$
(see Secs.~VI and VII of the SM~\cite{SM_ref}).
The microscopic values of the coefficients follow from
the tight-binding model of Eq.~(\ref{eq:heff}):
$\alpha_1 = \alpha\lambda J_{sf}a$, $\beta_1 = c_1 J_{sf}a^3$, $\gamma_1 = c_2 J_{sf}a^3$,
and $\delta = \gamma J_{sf}a^2$,
with $\alpha_1,\beta_1 \propto \lambda$ and $\gamma_1$ finite at $\lambda = 0$
(see Sec.~V of the SM~\cite{SM_ref}).

The resulting energy dispersion of the spin-split bands is given by $\varepsilon_{\pm}(\mathbf{q}) = \varepsilon_0(\mathbf{q}) \pm \sqrt{h_x^2 + h_y^2 + h_z^2}$, and the corresponding spin expectation values are $\langle s_i \rangle_{\pm} = \pm h_i(\mathbf{q}) / |\mathbf{h}(\mathbf{q})|$.
This expansion captures the evolution of the in-plane spin texture with $|\mathbf{q}|$.
For small $|\mathbf{q}|$ (inner Fermi surface)
the linear terms dominate, $\mathbf{h}_\parallel \approx \alpha_1(q_x, -q_y)$,
yielding the $p$-wave-like anti-vortex.
For larger $|\mathbf{q}|$ (outer Fermi surface)
the cubic terms take over,
$h_x \approx q_x(\beta_1 q_x^2 + \gamma_1 q_y^2)$
and $h_y \approx -q_y(\beta_1 q_y^2 + \gamma_1 q_x^2)$,
which manifest as the $f$-wave-like texture.
The out-of-plane component $h_z = \delta\,q_x q_y$
is dictated by the quadratic term alone
and preserves its even-parity $d$-wave symmetry throughout momentum space.

\emph{Appendix D: Edelstein response tensors}---
The non-relativistic Edelstein tensor $\chi_{ij}$,
defined by the current-induced spin accumulation $\langle\hat s_i\rangle=\chi_{ij}E_j$,
is evaluated within the Kubo formalism and
decomposed into an intraband ($T$-even) and an interband ($T$-odd) part
\cite{Li15,Hu25,Pari25}.
With $D_{n\mathbf k} \equiv (E_F - \epsilon_{n\mathbf{k}})^2 + \Gamma^2$,
\begin{align}
\chi^{\mathrm{intra}}_{ij} &=
\frac{e\hbar}{\pi V} \sum_{\mathbf{k},m,n}
\frac{\Gamma^2 \operatorname{Re}
\left(
\langle n\mathbf{k}|\hat{s}_i|m\mathbf{k}\rangle
\langle m\mathbf{k}|\hat{v}_j|n\mathbf{k}\rangle
\right)
}{D_{n\mathbf{k}}D_{m\mathbf{k}}},
\label{chi_intra}\\[4pt]
\chi^{\mathrm{inter}}_{ij}
&=
\frac{2e\hbar}{\pi V}
\sum_{\mathbf{k}}
\sum_{\substack{n \in \mathrm{occ} \\ m \in \mathrm{unocc}}}
\frac{
\operatorname{Im}
\left(
\langle n\mathbf{k}|\hat{s}_i|m\mathbf{k}\rangle
\langle m\mathbf{k}|\hat{v}_j|n\mathbf{k}\rangle
\right)
}{
(\epsilon_{n\mathbf{k}}-\epsilon_{m\mathbf{k}})^2
}.
\label{chi_inter}
\end{align}
Here $\Gamma$ represents the phenomenological scattering rate.
The terms $|n\mathbf{k}\rangle$ and $\epsilon_{n\mathbf{k}}$ denote the Bloch state
and its corresponding energy eigenvalue for the $n$-th band, with $E_F$ being the Fermi energy.
Finally, $\hat{s}_i$ and $\hat{v}_j$ are the $i$-th component of the spin operator
and the $j$-th component of the velocity operator, respectively.

The spin-group symmetries of $\mathrm{TbB}_4$ strictly restrict
the non-zero elements of the Edelstein tensor $\chi_{ij}$.
Specifically, the operation $\{E\|M_{z}\}$ enforces the constraint $\chi_{zz} E_z = -\chi_{zz} E_z$,
which requires $\chi_{zz} = 0$.
The screw-rotation operation $\{C_{2y}\|C_{2y}|\tau_{(a/2+b/2)}\}$ imposes
$\chi_{yx} E_x = -\chi_{yx} E_x$ and $\chi_{yz} E_z = -\chi_{yz} E_z$,
yielding $\chi_{yx} = \chi_{yz} = 0$;
similarly, the operation $\{C_{2x}\|C_{2x}|\tau_{(a/2+b/2)}\}$ dictates $\chi_{xy} = \chi_{xz} = 0$.
Furthermore, the operation $\{I_{s}C_{4z}\|C_{4z}^{-1}\}$ relates the remaining tensor elements,
enforcing $\chi^{\text{intra}}_{xx} = -\chi^{\text{intra}}_{yy}$
and $\chi^{\text{inter}}_{xx} = \chi^{\text{inter}}_{yy}$ \cite{spinsym}.
Consequently, the only symmetry-allowed components are $\chi_{xx}$ and $\chi_{yy}$.

\emph{Appendix E: Spin Hall conductivity and spin-group symmetry}---
The spin Hall conductivity $\sigma^i_{jk}$,
defined by $J^i_j = \sigma^i_{jk} E_k$
with the spin current operator $\hat J^i_j \equiv \tfrac12\{\hat s_i,\hat v_j\}$,
consists of a $T$-even (intrinsic) and a $T$-odd (spin-splitter) contribution.
The former is given by the spin Berry curvature~\cite{Guo08,Freimuth14,Gonzalenz21,Hu25},
\begin{align}
\sigma^{i,{\rm even}}_{jk} &= -\frac{2e\hbar}{V} \sum_{\mathbf{k}} \Omega^i_{jk}(\mathbf{k}), \nonumber \\
\Omega^i_{jk}(\mathbf{k}) &= \sum_{\substack{n \in \mathrm{occ} \\ m \in \mathrm{unocc}}}
\frac{ \operatorname{Im} \left( \langle n\mathbf{k}| \hat J^i_j |m\mathbf{k}\rangle \langle m\mathbf{k}|\hat{v}_k|n\mathbf{k}\rangle \right) }{(\epsilon_{n\mathbf{k}}-\epsilon_{m\mathbf{k}})^2},
\label{eq:she_even}
\end{align}
and the latter is the Fermi-surface counterpart,
\begin{equation}
\sigma^{i,{\rm odd}}_{jk}=-\frac{e\hbar}{\pi V}\sum_{\mathbf k,m,n}
\frac{\Gamma^2\,{\rm Re}\big[\langle n\mathbf k| \hat J^i_j |m\mathbf k\rangle
\langle m\mathbf k|\hat v_k|n\mathbf k\rangle\big]}
{D_{n\mathbf k}D_{m\mathbf k}},
\label{eq:she_odd}
\end{equation}
where $e$ is the electron charge and $\hbar$ is the reduced Planck constant.
The other symbols are as in Appendix~D.
The integrand $s_i v_j v_k$ of Eq.~(\ref{eq:she_odd}) is odd under the time-reversal operator,
whereas that of Eq.~(\ref{eq:she_even}) is even,
the imaginary part of the matrix-element product supplying an additional sign
under the time-reversal operator.
Accordingly, the $T$-odd and $T$-even contributions reverse and remain unchanged, respectively,
upon time reversal of the magnetic order.

Due to the $\{C_{2z}\|P\}$ symmetry
(which is common to $\mathrm{GdB}_4$, $\mathrm{TmB}_4$, and $\mathrm{TbB}_4$),
the following relations hold:
$v_i(\mathbf{k}) = -v_i(-\mathbf{k})$, $s_{x,y}(\mathbf{k}) = -s_{x,y}(-\mathbf{k})$,
and $s_z(\mathbf{k}) = s_z(-\mathbf{k})$.
Therefore, strict constraints are imposed on the spin Berry curvature,
$\Omega^{x,y}_{jk}(\mathbf k) = -\Omega^{x,y}_{jk}(-\mathbf k)$
and $\Omega^z_{jk}(\mathbf k) = \Omega^z_{jk}(-\mathbf k)$,
and likewise on the integrand of Eq.~(\ref{eq:she_odd}),
which has the same spin and velocity structure.
Consequently, all in-plane components vanish for both $T$-even and $T$-odd contributions
($\sigma^x_{jk} = \sigma^y_{jk} = 0$).
Indeed, although the local spin Berry curvature is nonzero
for both $\Omega^{x}_{xy}$ and $\Omega^{z}_{xy}$ (see Sec.~XII of the SM~\cite{SM_ref}),
only $\Omega^{z}_{xy}$ survives the Brillouin-zone integration in $\mathrm{TbB}_4$.
The element $\{E\|M_z\}$ further enforces
$\sigma^z_{zx}=\sigma^z_{xz}=\sigma^z_{zy}=\sigma^z_{yz}=0$ for both contributions,
leaving only $\sigma^z_{xy}$ and $\sigma^z_{yx}$,
which in $\mathrm{TbB}_4$ are related by $\{I_sC_{4z}\|C_{4z}^{-1}\}$ as
$\sigma^{z,{\rm even}}_{xy} = -\sigma^{z,{\rm even}}_{yx}$
and $\sigma^{z,{\rm odd}}_{xy} = \sigma^{z,{\rm odd}}_{yx}$.
The two contributions differ among the compounds.
In $\mathrm{GdB}_4$ the $PT$ symmetry makes every band spin-degenerate,
so the $T$-odd contribution vanishes while the $T$-even one remains finite.
In $\mathrm{TmB}_4$ the element $\{I_s\|C_{2x}|\tau_{(a/2+b/2)}\}$ enforces
$\sigma^{z,{\rm even}}_{xy} = \sigma^{z,{\rm even}}_{yx}=0$,
whereas the $T$-odd contribution, which acquires an additional sign from $I_s$,
is allowed, as expected for a $d$-wave altermagnet.
In $\mathrm{TbB}_4$ both contributions are finite
(Figs.~\ref{edel}(c) and (d)).

\clearpage
\onecolumngrid
\includepdf[pages=1]{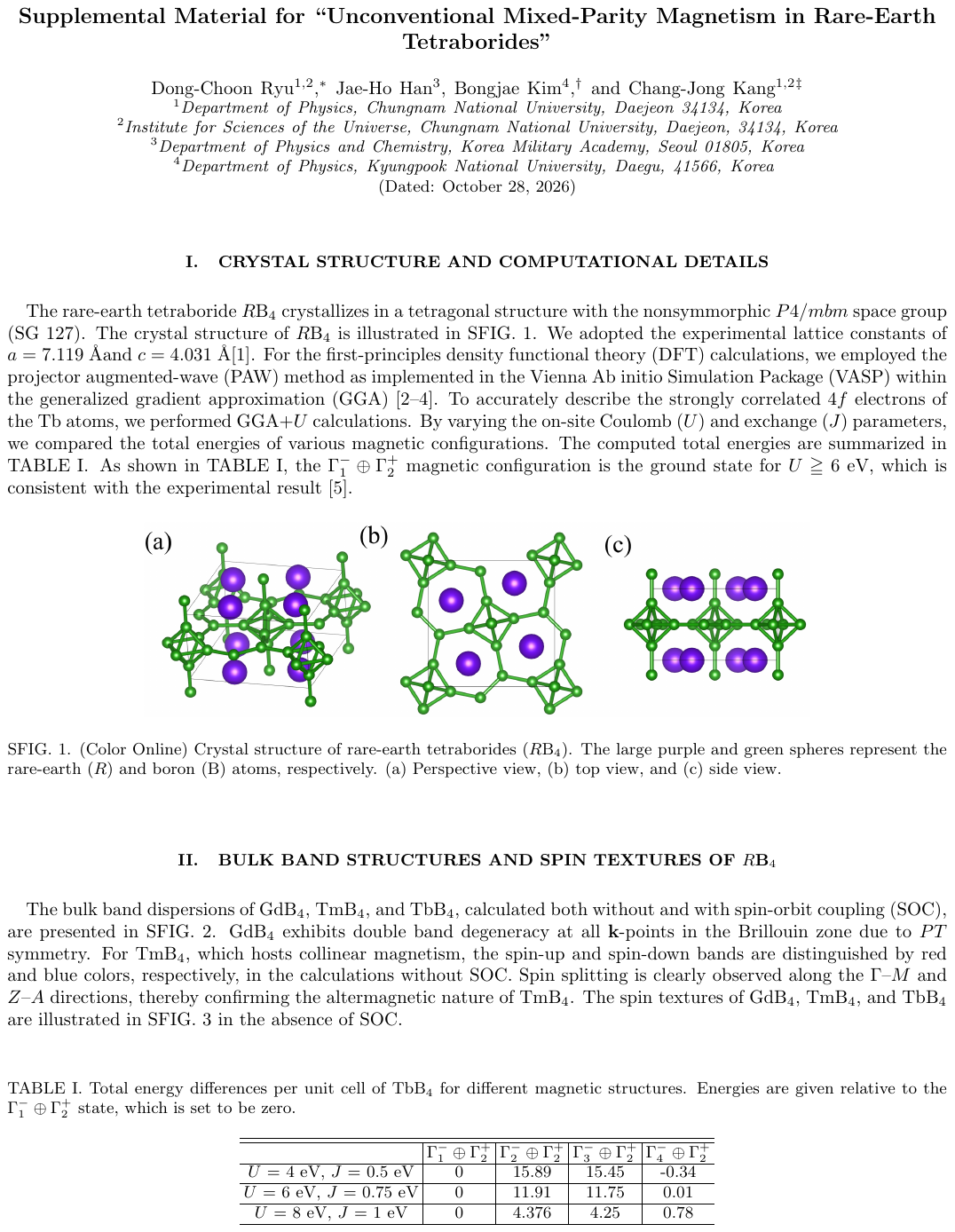}

\clearpage
\includepdf[pages=2]{TbB4_suppl.pdf}

\clearpage
\includepdf[pages=3]{TbB4_suppl.pdf}

\clearpage
\includepdf[pages=4]{TbB4_suppl.pdf}

\clearpage
\includepdf[pages=5]{TbB4_suppl.pdf}

\clearpage
\includepdf[pages=6]{TbB4_suppl.pdf}

\clearpage
\includepdf[pages=7]{TbB4_suppl.pdf}

\clearpage
\includepdf[pages=8]{TbB4_suppl.pdf}

\clearpage
\includepdf[pages=9]{TbB4_suppl.pdf}

\clearpage
\includepdf[pages=10]{TbB4_suppl.pdf}

\clearpage
\includepdf[pages=11]{TbB4_suppl.pdf}

\clearpage
\includepdf[pages=12]{TbB4_suppl.pdf}

\clearpage
\includepdf[pages=13]{TbB4_suppl.pdf}

\clearpage
\includepdf[pages=14]{TbB4_suppl.pdf}

\clearpage
\includepdf[pages=15]{TbB4_suppl.pdf}

\clearpage
\includepdf[pages=16]{TbB4_suppl.pdf}

\clearpage
\includepdf[pages=17]{TbB4_suppl.pdf}

\clearpage
\includepdf[pages=18]{TbB4_suppl.pdf}

\clearpage
\includepdf[pages=19]{TbB4_suppl.pdf}

\clearpage
\includepdf[pages=20]{TbB4_suppl.pdf}
\twocolumngrid

\end{document}